# SCMA Codebook Design

Mahmoud Taherzadeh, Hosein Nikopour, Alireza Bayesteh, and Hadi Baligh
Ottawa Wireless R&D Centre
Huawei Technologies Canada Co., LTD.
Ottawa, Ontario, Canada
{mahmoud.taherzadeh, hosein.nikopour, alireza.bayesteh, hadi.baligh}@huawei.com

*Abstract*—Multicarrier CDMA is a multiple access scheme in which modulated QAM symbols are spread over OFDMA tones by using a generally complex spreading sequence. Effectively, a QAM symbol is repeated over multiple tones. Low density signature (LDS) is a version of CDMA with low density spreading sequences allowing us to take advantage of a near optimal message passing algorithm (MPA) receiver with practically feasible complexity. Sparse code multiple access (SCMA) is a multi-dimensional codebook-based non-orthogonal spreading technique. In SCMA, the procedure of bit to QAM symbol mapping and spreading are combined together and incoming bits are directly mapped to multi-dimensional codewords of SCMA codebook sets. Each layer has its dedicated codebook. Shaping gain of a multi-dimensional constellation is one of the main sources of the performance improvement in comparison to the simple repetition of QAM symbols in LDS. Meanwhile, like LDS, SCMA enjoys the low complexity reception techniques due to the sparsity of SCMA codewords. In this paper a systematic approach is proposed to design SCMA codebooks mainly based on the design principles of lattice constellations. Simulation results are presented to show the performance gain of SCMA compared to LDS and OFDMA.

## I. INTRODUCTION

Future 5G wireless networks are expected to support very diverse traffics with much tighter requirements [1]. Massive connectivity, better quality of service, higher throughput, lower latency, or lower control signaling overhead are some of the requirements that should be met with new waveform and access designs. Sparse code multiple access (SCMA) [2] is a multi-dimensional codebook-based non-orthogonal spreading technique which can be seen as a generalization of low density signature (LDS) [3] to address the above requirements. LDS is a special approach of CDMA spreading with a few numbers of nonzero elements within a large signature length. The low density characteristic of LDS signatures lets us to take advantage of the low complexity message passing algorithm (MPA) detector with ML-like performance. MPA performs well even if the system is overloaded with a number of non-orthogonal layers larger than the spreading factor.

In SCMA, incoming bits are directly mapped to multi-dimensional complex codewords selected from a predefined codebook set. In other words, in SCMA the QAM mapper and the CDMA (or LDS) spreader are merged together to directly map a set of bits to a complex sparse vector so called a *codeword*. Every layer has a specific SCMA codebook set. The overlaid layers are non-orthogonally super-imposed on top of each other. Same as LDS, codewords of SCMA are sparse such that the MPA detection technique is applicable with a moderate complexity. In addition, the system can be overloaded where the number of multiplexed layers is more than the spreading factor (or equivalently the length of a codeword).

SCMA replaces QAM modulation and LDS spreading with multi-dimensional codebooks. This enables SCMA to benefit from shaping or coding gains [4] of multi-dimensional constellations as opposed to simple repetition code of LDS. Hence, SCMA improves the spectral efficiency of LDS through multi-dimensional shaping gain of codebooks while it still provides the benefits of LDS in terms of overloading and moderate complexity of detection.

This paper proposes a systematic approach toward the design of SCMA codebooks. Multi-dimensional constellation design is studied in different aspects of communications [4],[5],[6]. The SCMA codebook design is even more complicated as multiple layers are multiplexed with different codebooks. Following the proposed design procedure, first a multi-dimensional constellation is designed with a good Euclidean distance profile. The base constellation is then rotated to achieve a reasonable product distance. The approach follows the principles of code design for point-to-point communication over fast fading channel [5],[7],[8]. Having the rotated constellation, different sets of operators such as phase rotations are applied on top of it to build multiple sparse codebooks for several layers of SCMA.

The rest of the paper is organized as follows. Section II defines the system model and SCMA system structure. Section III is dedicated to the multi-dimensional codebook design for SCMA. This section describes the procedure of the proposed multi-stage SCMA codebook design. Numerical results are reported in Section IV comparing SCMA with LDS and OFDMA. The paper finally concludes in Section V.

## II. SYSTEM MODEL AND DESCRIPTION

An SCMA encoder is defined as a map from $\log_2(M)$ bits to a $K$-dimensional complex codebook of size $M$. The $K$-dimensional complex codewords of the codebook are sparse vectors with $N < K$ non-zero entries. All the codewords in the codebook contain 0 in the same $K - N$ dimensions. A mapping matrix $\mathbf{V}$ maps the $N$ non-zero dimensions to the $K$-dimensional complex domain. Similar to optical orthogonal codes (OOC) [9], this can be also represented by a binary vector $\mathbf{f}$ of length $K$ indicating the positions of nonzero entries of the codebook.

An SCMA encoder contains $J$ separate layers. The constellation function $g_j$ generates the constellation set $\mathcal{C}_j$ with

$M_j$ alphabets of length $N_j$. The mapping matrix $\mathbf{V}_j$ maps the $N_j$-dimensional constellation points to SCMA codewords to form the codeword set $\mathcal{X}_j$. Without loss of generality, here we assume that all layers have the same constellation size and length, i.e. $M_j = M, N_j = N, \forall j$. SCMA codewords are multiplexed over $K$ shared orthogonal resources, e.g. OFDMA tones. The received signal after the synchronous layer multiplexing can be expressed as $\mathbf{y} = \sum_{j=1}^{J} \text{diag}(\mathbf{h}_j)\mathbf{x}_j + \mathbf{n}$ where $\mathbf{x}_j = \mathbf{V}_j g_j(\mathbf{b}_j)$ is the vector of SCMA codeword of layer $j$, $\mathbf{h}_j$ is the channel vector of layer $j$ and $\mathbf{n}$ is the ambient noise. In the case that all layers are transmitted from the same transmit point, all the channels to a target receiver are identical, i.e. $\mathbf{h}_j = \mathbf{h}, \forall j$. By multiplexing $J$ layers over $K$ resources, the overloading factor of the code is defined as $\lambda := J/K$.

The received signal at resource $k$ is presented by $y_k$. As the codewords $\mathbf{x}_j$'s are sparse, only a few of them collide over resource $k$. The set of resources occupied by layer $j$ depends on the mapping matrix $\mathbf{V}_j$ and the set is determined by the index of the non-zero elements of binary indicator vector $\mathbf{f}_j$ corresponding to the non-zero rows of $\mathbf{V}_j$. The whole structure of SCMA code can be represented by a factor graph matrix defined as $\mathbf{F} = (\mathbf{f}_1, \ldots, \mathbf{f}_J)$. Layer node $j$ and resource node $k$ are connected if and only if $(\mathbf{F})_{kj} = 1$. An example of a factor graph representation of $\mathbf{F}$ is shown in Fig. 2 with 6 symbol nodes and 4 resource nodes.

Given the received signal $\mathbf{y}$ and channel knowledge $\{\mathbf{h}_j\}_{j=1}^{J}$, the near optimal detection of $J$ layers can be preformed iteratively by applying the MPA detector over the underlying factor graph. The complexity of MPA is proportional to $M^{d_f}$ where $d_f$ is the number branches arriving to a resource node. Sparsity pattern of the codebooks helps to limit the number of braches per resource node as hence the complexity of the MPA receiver.

### III. SCMA CODEBOOK DESIGN

The design problem of an SCMA code with structure $\mathcal{S}(\mathcal{V}, \mathcal{G}; J, M, N, K)$, $\mathcal{V} := [\mathbf{V}_j]_{j=1}^{J}$ and $\mathcal{G} := [g_j]_{j=1}^{J}$ can be defined as

$$\mathcal{V}^*, \mathcal{G}^* = \arg\max_{\mathcal{V},\mathcal{G}} m(\mathcal{S}(\mathcal{V}, \mathcal{G}; J, M, N, K)) \quad (1)$$

where $m$ is a given design criterion. As the appropriate definition of $m$ and solution of this multi-dimensional problem is unknown, a multi-stage optimization approach is proposed to achieve a sub-optimal solution for the problem.

#### A. Mapping Matrix

As described before, the set of mapping matrices $\mathcal{V}$ determines the number of layers interfering at each resource node which in turn defines the complexity of the MPA detection. The sparser the codewords the less complex is the MPA detection. When the maximum overloading is desired, the unique solution of $\mathcal{V}$ is simply determined by inserting $K - N$ all-zero row vectors within the rows of $\mathbf{I}_N$. The solution holds the following properties: $J = \binom{K}{N}$, $d_f = \binom{K-1}{N-1} = \frac{JN}{K}$, $\forall j$, $\lambda = \frac{J}{K} = \frac{d_f}{N}$, and $\max(0, 2N - K) \leq l \leq N - 1$, where $l$ is the number of the overlapping elements of any two distinct $\mathbf{f}_j$ vectors.

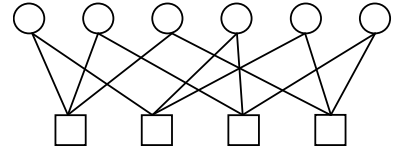

Fig. 1. Factor graph representation of an SCMA system with $J = 6$, $K = 4$, $N = 2$, and $d_f = 3$.

#### B. Constellation Points and Multi-dimensional Mother Constellation

Having the mapping set $\mathcal{V}^+$, the optimization problem of an SCMA is reduced to

$$\mathcal{G}^+ = \arg\max_{\mathcal{G}} m(\mathcal{S}(\mathcal{V}^+, \mathcal{G}; J, M, N, K)). \quad (2)$$

The problem is to define $J$ different $N$-dimensional constellations each containing $M$ points. To simplify the optimization problem, the constellation points of the layers are modeled based on a mother constellation and layer-specific operators, i.e. $g_j \equiv (\Delta_j)g, \forall j$, where $\Delta_j$ denotes a *constellation operator*. According to the model, the code optimization problem turns into

$$g^+, [\Delta_j^+]_{j=1}^{J} =$$
$$\arg\max_{g,[\Delta_j]_{j=1}^{J}} m\left(\mathcal{S}(\mathcal{V}^+, \mathcal{G} = [(\Delta_j)g]_{j=1}^{J}; J, M, N, K)\right). \quad (3)$$

As a sub-optimal approach to the above problem, the mother constellation and the operators are determined separately.

*1) Design Metrics and Rotated Contellations*

Large minimum Euclidean distance of a multi-dimensional constellation ensures a good performance of the SCMA system with a small number of layers where there are no collisions between the layers over a tone. Once the number of layers grows, two or more layers may collide over a tone. Under this condition, it is important to induce dependency among the non-zero elements of codewords to be able to recover colliding codewords from the other tones. In addition, power imbalance across the dimensions of codewords introduces near-far effect among colliding layers. It helps MPA detector to operate more effectively to remove interferences among co-paired layers.

Having a constellation with a desirable Euclidian distance profile, a unitary rotation can be applied on the base constellation to control dimensional dependency and power variation of the constellation while maintaining the Euclidian distance profile unchanged. Inspiring from code design for communications over fast fading channel [7],[8], a unitary rotation might be designed to maximize the minimum product distance of the constellation. Therefore, the design objective encapsulates both the sum distance and the product distance between the points in the mother constellation. Similar to communication over fading channel, product distance becomes the dominant performance indicator for high SNR ranges.

*2) Rotated Lattice Constellations*

In general, the base constellation can be any multi-dimensional constellation with a maximized minimum Euclidean distance. For low rates, constellation design can be done by heuristic optimization, but for higher rates a structured construction approach is required. Lattice constellation is a structural approach of the base constellation design. As a special case of lattice constellations, we can consider the base constellation to be formed by orthogonal QAMs on different complex planes. It is equivalent to a constellation from the lattice $\mathbb{Z}^{2N}$. Gray labeling is an advantage of this type of lattice constellations. Unitary rotations of QAM lattice constellations are optimized in [7] for dimensions 2 to 4 in order to maximize the minimum product distance of rotated lattices.

*3) Shuffling Multi-dimensional Constellations in Real and Imaginary Axes*

If a complex constellation is built such that its real part is independent of its imaginary part, it can help to reduce the decoding complexity while yet maintaining dependency among the complex dimensions of the resulted multi-dimensional constellation. Using this technique, the complexity order of MPA reduces from $M^{d_f}$ to $M^{d_f/2}$ which is a high complexity saving especially for large constellation sizes.

A shuffling method is proposed in this section as illustrated in Fig. 2 to separate real and imaginary parts. The idea is to construct an $N$-dimensional complex mother constellation from Cartesian product of two $N$-dimensional real constellations, where each of them is constructed by the same method described in the previous section. One of these two $N$-dimensional real constellations corresponds to the real part of the points of the complex mother constellation and the other one corresponds to the imaginary part.

Fig. 3 shows an example of the shuffling to construct a 16-point SCMA mother constellation (T16QAM) applicable to codebooks with two nonzero positions ($N$=2). The optimum rotation angle that maximizes the minimum product distance is $\tan^{-1}\left(\frac{1+\sqrt{5}}{2}\right)$.

*4) Rotation to Minimize Number of Projection Points*

For the sake of simplicity of the MPA decoding, it is more desirable to use mother constellations that have a smaller number of projections per tone (or complex dimension). Let $m$ denote the number of projects per complex dimensions of an $M$-point constellation. It is obvious that $m \leq M$. As $m$ decreases, the complexity of the corresponding MPA detector is also reduced by $m^{d_f}$. During the process of mother constellation design, the rotation matrix can be set in a way that it leads to the lower number of projected points. It makes the minimum product distance equal to zero and degrades the high SNR performance of the SCMA system. Consequently, there is a trade-off between the high SNR performance and complexity in this case. As an example, Fig. 4 shows a solution with 9 projections per complex dimension of a 16-point constellation. According to simulations result, it performs close to T16QAM for mid-SNR range while it reduces complexity order from $16^3$ to $9^3$.

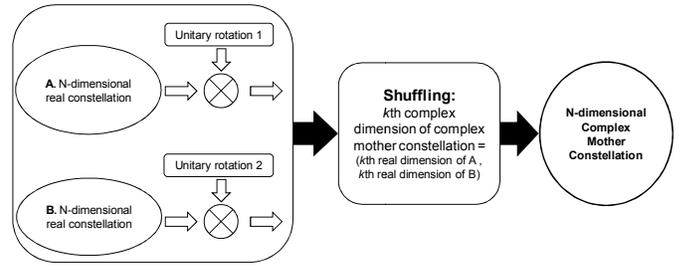

Fig. 2. Shuffling of real and imaginary axes to mix two orthogonal sub-constellations.

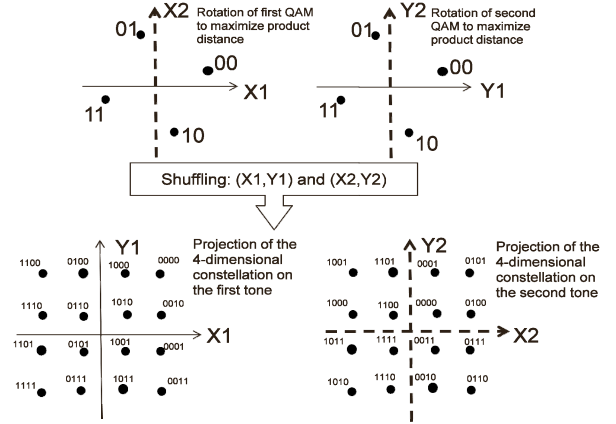

Fig. 3. An example of a 16-point SCMA constellation for SCMA codewords with two nonzero elements.

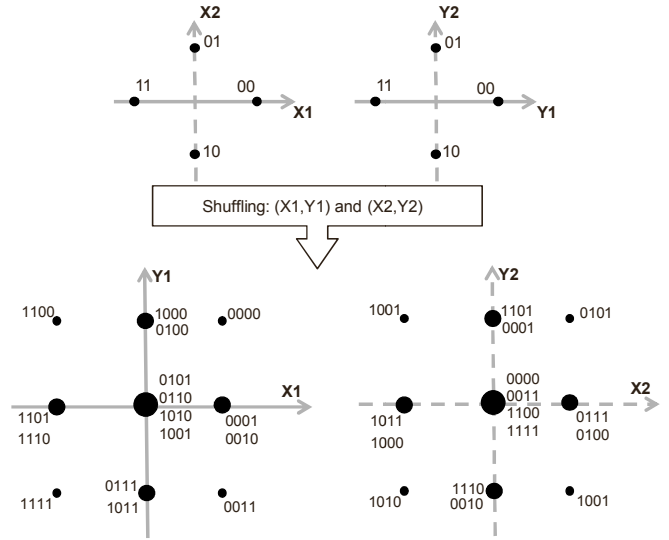

Fig. 4. 16-point SCMA constellation with 9 projection points per complex dimension.

*C. Constellation Function Operators*

By having a solution for the mother constellation ($\mathcal{C}^+$ or equivalently $g^+$), the original SCMA optimization problem is further reduced to optimization of layer specific operators, i.e.

$$[\Delta_j^+]_{j=1}^J = \arg\max_{[\Delta_j]_{j=1}^J} m\left(\mathcal{S}(\mathcal{V}^+, \mathcal{G} = [(\Delta_j)g^+]_{j=1}^J; J, M, N, K)\right). \quad (4)$$

The operators may include phase rotation, and layer power offset. The codebooks of different SCMA layers are constructed based on the mother constellation $g$ and a layer-specific operator $\Delta_j$ for layer $j$. In uplink, as the layers pass through different channels, layer specific phase rotation loses its importance. However, in downlink for layers with same channel experience, the destructive codeword collision can be avoided by carful design of phase rotations and power levels of different layers. As a result, the points of the overall super-constellation of the combined layers become distinct. In this paper, LDS-like phase signatures [10] are used to build multiple SCMA codebooks on top of the mother constellation with no power offsets.

IV. NUMERICAL RESULTS

In the following, the link-level performance of SCMA is studied and the performance gain of SCMA over LDS and OFDMA is established.

A. Dimensional Power Variation of SCMA Codewords

One feature of SCMA is that, unlike LDS, the transmitted signals over the non-zero tones of each codeword have different powers. This power variation, which is the benefit of rotated lattice constellation, helps MPA to operate more efficiently to cancel inter-layer interferences. To see this effect, SCMA and LDS are compared in Fig. 5(a) for 2, 4, and 6 layers and the MCS per layer assumed to be 4-point constellation with turbo code rate of 0.75. The turbo code follows the structure of the long-term evolution (LTE) standard [11]. Both SCMA and LDS follow the factor graph of Fig. 1 with basic parameters of $K=4$, $J=6$, and $d_f=3$. The structure of the signature matrix of LDS is very similar to the one reported in [10]. All simulations are performed in AWGN channel. The horizontal axis of Fig. 5(a) is the SNR per layer to reflect the degradation with respect to the single layer performance. LDS and SCMA show very similar performance for 2 non-overlapped layers. However, as the number of layers increases, SCMA starts to outperform LDS due to its stronger inter-layer interference cancellation capability. Note that with the selected 4-point SCMA codebooks, the multi-dimensional shaping gain is negligible. That is why SCMA and LDS with 1 or 2 non-overlapped layers perform identically. Therefore, the gain illustrated here for larger number of layers is purely due to the dimensional power variation of SCMA codebooks rather than the shaping gain.

B. Constellation Shaping Gain

Another advantage of SCMA over LDS is its inherent shaping gain due to the fact that SCMA enjoys additional degrees of freedom in the multi-dimensional constellations design. Note that this feature can be observed better with lower number of layers (1 or 2) where there is no inter-layer interference and the layers are totally separated with no overlaps. As shown in Fig. 5(b), 16-point SCMA codebooks and 16QAM LDS constellation are compared with 2 layers and turbo code rate of 0.5 with overall rate of 1 bit/tone. The horizontal axis represent the total receive signal power to noise ratio. The simulation results confirm that SCMA provides huge performance gain over 16QAM LDS in AWGN channel due to the shaping gain of the SCMA multi-dimensional codebooks. It also illustrates the weakness of LDS repetition coding specially for larger constellation sizes such as 16QAM. Note that SCMA outperforms LDS in this scenario even if LDS operates with the best combination of MCS and number of layers (QPSK with 6 layers and code rate 1/3) for the given total rate of 1 bit/tone.

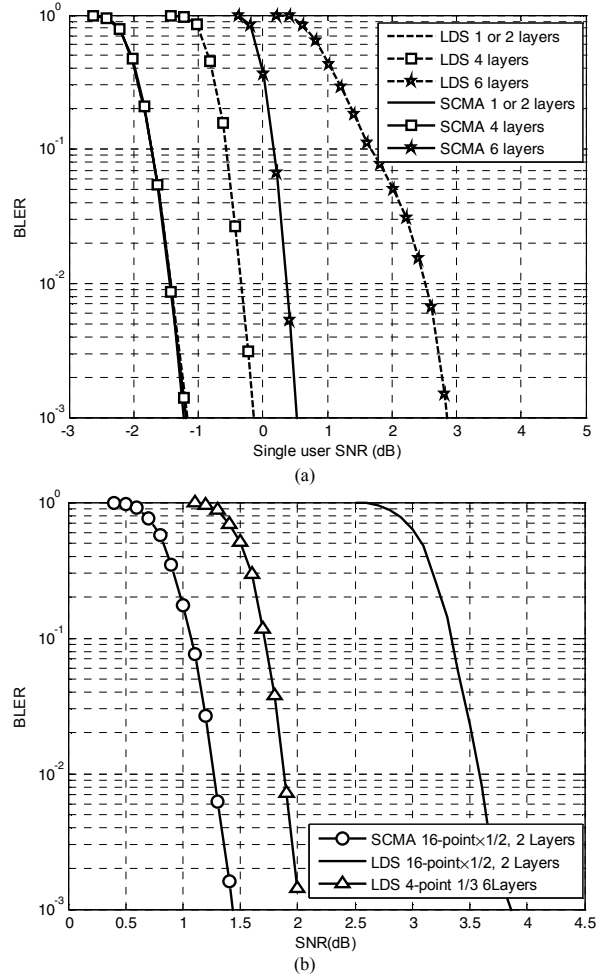

Fig. 5. SCMA vs. LDS: (a) co-paired interference cancellation capability of SCMA vs. LDS due to dimensional power variation of SCMA codewords, (b) shaping gain of multi-dimensional 16-ponit SCMA codebook (T16QAM) vs. LDS repetitive 16QAM constellation.

C. Performance in Uplink Fading Channels

Performance of SCMA is evaluated through link-level simulation for uplink transmission in fading channel. Every SCMA layer is carried over OFDMA tones in a pedestrian B (PB) fading channel with speed of 3 km/h. The carrier frequency is 2.6 GHz and the frequency spacing of the OFDMA tones is 15 kHz as in LTE standard. A data payload occupies 24 LTE resource blocks (RBs). Antenna configuration is one transmit and two uncorrelated receive antennas.

Fig. 6 illustrates the BLER performance comparing SCMA, LDS, OFDMA, and LTE single carrier FDMA (SC-FDMA) waveforms, respectively. The overall spectral efficiency (SE) is set to 1.5 bits/tone. For each case, the best combination in terms of the number of layers, modulation, and coding rate is selected to achieve the best performance for the given SE. The total transmit power is assumed to be the same for all waveforms. As can be observed in the figure, SCMA outperforms LDS, OFDMA, and SC-FDMA and the gain is over 2 dB compared to OFDMA and SC-FDMA.

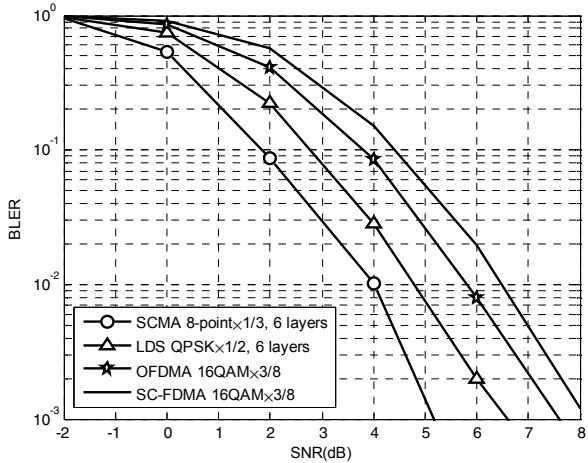

Fig. 6. Performance comparison between SCMA, LDS, OFDMA, and SC-FDMA in uplink with SIMO fading channel.

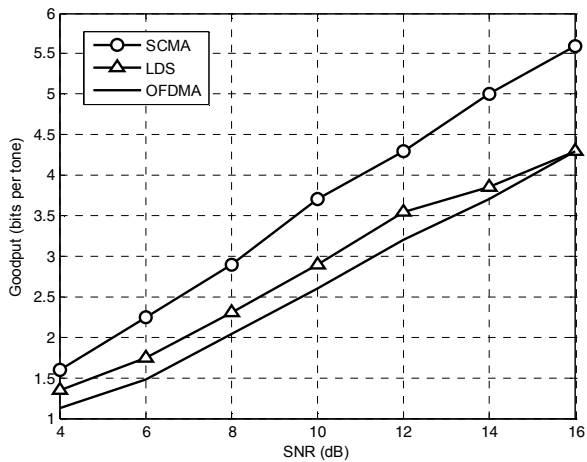

Fig. 7. Uplink goodput of SCMA vs. OFDMA and LDS over different ranges of SNR with SIMO fading channel.

The comparison of uplink goodput of SCMA, LDS and OFDM is shown in Fig. 7 for a wide range of SNR. For each given SNR, the appropriate MCS and number of layers are selected to guarantee the best performance possible for each waveform. The gain of SCMA is obvious and it grows as the SNR increases.

## V. CONCLUSION

A systematic multi-stage method is proposed for codebook design of SCMA. Lattice rotation technique is applied to design a multi-dimensional mother constellation. A base lattice constellation with a desired Euclidian distance profile is rotated to induce dimensional dependency and power variation while maintaining the Euclidean distance profile unchanged. Afterward, layer specific operators are applied on the mother constellation to build a codebook for every layer of SCMA. Simulation results illustrate the gain of SCMA over LDS and OFDMA for both AWGN and fading channels. Conclusively, SCMA has all the multi-access benefits of LDS in terms of overloading, moderate complexity, interference whitening, etc while it avoids the poor link performance of LDS. With these advantages, SCMA can potentially result in a good system performance in both uplink and downlink multiple access scenarios of future wireless networks.

ACKNOWLEDGEMENT

Part of this work has been performed in the framework of the FP7 project ICT-317669 METIS. The authors would like to acknowledge the contributions of their colleagues in METIS for the discussions and the comments.